\documentclass[10pt,twocolumn,pra,aps,superscriptaddress,nofootinbib,floatfix,showpacs]{revtex4}
\usepackage[latin9]{inputenc}
\usepackage{amsmath}
\usepackage{amssymb}
\usepackage{graphicx}
\usepackage{esint}

\makeatletter
\@ifundefined{textcolor}{}
{%
 \definecolor{BLACK}{gray}{0}
 \definecolor{WHITE}{gray}{1}
 \definecolor{RED}{rgb}{1,0,0}
 \definecolor{GREEN}{rgb}{0,1,0}
 \definecolor{BLUE}{rgb}{0,0,1}
 \definecolor{CYAN}{cmyk}{1,0,0,0}
 \definecolor{MAGENTA}{cmyk}{0,1,0,0}
 \definecolor{YELLOW}{cmyk}{0,0,1,0}
}

\usepackage{subfigure}

\usepackage[all]{xy}\usepackage{appendix}\usepackage{enumerate}\usepackage{color}

\usepackage{subfigure}

\makeatother

\begin{document}

\title{Sudden and slow quenches into the antiferromagnetic phase
of ultracold fermions}

\author{M.\,Ojekhile}

\affiliation{Zentrum f\"{u}r Optische Quantentechnologien and Institut f\"{u}r Laserphysik,
Universit\"{a}t Hamburg, 22761 Hamburg, Germany}

\affiliation{Institut f\"{u}r Laserphysik, Universit\"{a}t Hamburg, 22761 Hamburg, Germany}

\author{R.\,H\"{o}ppner}

\affiliation{Zentrum f\"{u}r Optische Quantentechnologien and Institut f\"{u}r Laserphysik,
Universit\"{a}t Hamburg, 22761 Hamburg, Germany}

\affiliation{Institut f\"{u}r Laserphysik, Universit\"{a}t Hamburg, 22761 Hamburg, Germany}

\author{H.\,Moritz}

\affiliation{Institut f\"{u}r Laserphysik, Universit\"{a}t Hamburg, 22761 Hamburg, Germany}

\author{L.\,Mathey}
\email{ludwig.mathey@physik.uni-hamburg.de}

\affiliation{Zentrum f\"{u}r Optische Quantentechnologien and Institut f\"{u}r Laserphysik,
Universit\"{a}t Hamburg, 22761 Hamburg, Germany}

\affiliation{Institut f\"{u}r Laserphysik, Universit\"{a}t Hamburg, 22761 Hamburg, Germany}
\begin{abstract}
We propose a method to reach the antiferromagnetic state of two-dimensional
Fermi gases trapped in optical lattices: Independent subsystems are
prepared in suitable initial states and then connected by a sudden
or slow quench of the tunneling between the subsystems. Examples of
suitable low-entropy subsystems are double wells or plaquettes, which
can be experimentally realized in Mott insulating shells using optical
super-lattices.
 We estimate the effective temperature $T^{*}$ of the system after
the quench by calculating the distribution of excitations created
using the spin wave approximation in a Heisenberg model. We investigate
the effect of an initial staggered magnetic field and find that for
an optimal polarization of the initial state the effective temperature
can be significantly reduced from $T^{*}\approx1.7\, T_{c}$ at zero
polarization to $T^{*}<0.65\, T_{c}$, where $T_{c}$ is the crossover
temperature to the antiferromagnetic state. 
 The temperature can be further reduced by using a finite quench time. 
We also show that $T^{*}$
decreases logarithmically with the linear size of the subsystem. 
\end{abstract}
\pacs{67.85.-d, 71.10.Fd, 75.30.Ds}
\maketitle
\begin{center}
\textbf{I. INTRODUCTION} 
\par\end{center}
Ultracold fermions trapped in optical lattices have emerged as ideal
model systems to simulate the Hubbard model \cite{Zwerger2008,Esslinger2010}.
Major achievements include the observation of Mott insulating states
\cite{Joerdens2008,Schneider2008} and of short-range antiferromagnetic
ordering \cite{Esslinger2013} and the simulation thereof \cite{Greiner2011}.
Currently the observation of long-range antiferromagnetic ordering
is the next milestone on the quest to simulate low-temperature phases
of the Hubbard model and find answers to open questions such as the
origin of high-temperature superconductivity. The main experimental
challenge is the reduction of entropy or temperature by at least a
factor of two or four, respectively \cite{Joerdens2010}.

To achieve this temperature reduction, one can take advantage of the
excellent tunability of ultracold atomic systems: Several cooling
schemes have been proposed, which should allow to reduce the entropy
of the systems (for a review, see Ref.  \cite{deMarco2011}). In the scheme
proposed here, we make use of the fact that in trapped systems, gapped
states can be in thermal contact with gapless states, similarly as
in related proposals \cite{Capogrosso2008,Bernier2009,Ho2009,Cirac2011,Trivedi2011,Hulet2012,Vekua2011}.
 As a result, the entropy in the gapped region is strongly reduced
and the excitations in the gapless region carry most of the entropy.
A prominent example is a band insulating state surrounded by a compressible
metallic state.

We propose to transform the central band insulating shell into an
antiferromagnetic state by first assembling subsystems featuring precursors
of magnetic ordering, followed by a sudden or slow quench connecting
the previously independent subsystems. We calculate the effective
temperature $T^{*}$ after the quench and compare it with the critical
temperature for antiferromagnetic ordering $T_{c}$. The following
questions then arise naturally: What is the ideal inital state that
minimizes $T^{*}$? 
\begin{figure}
\includegraphics[scale=0.33]{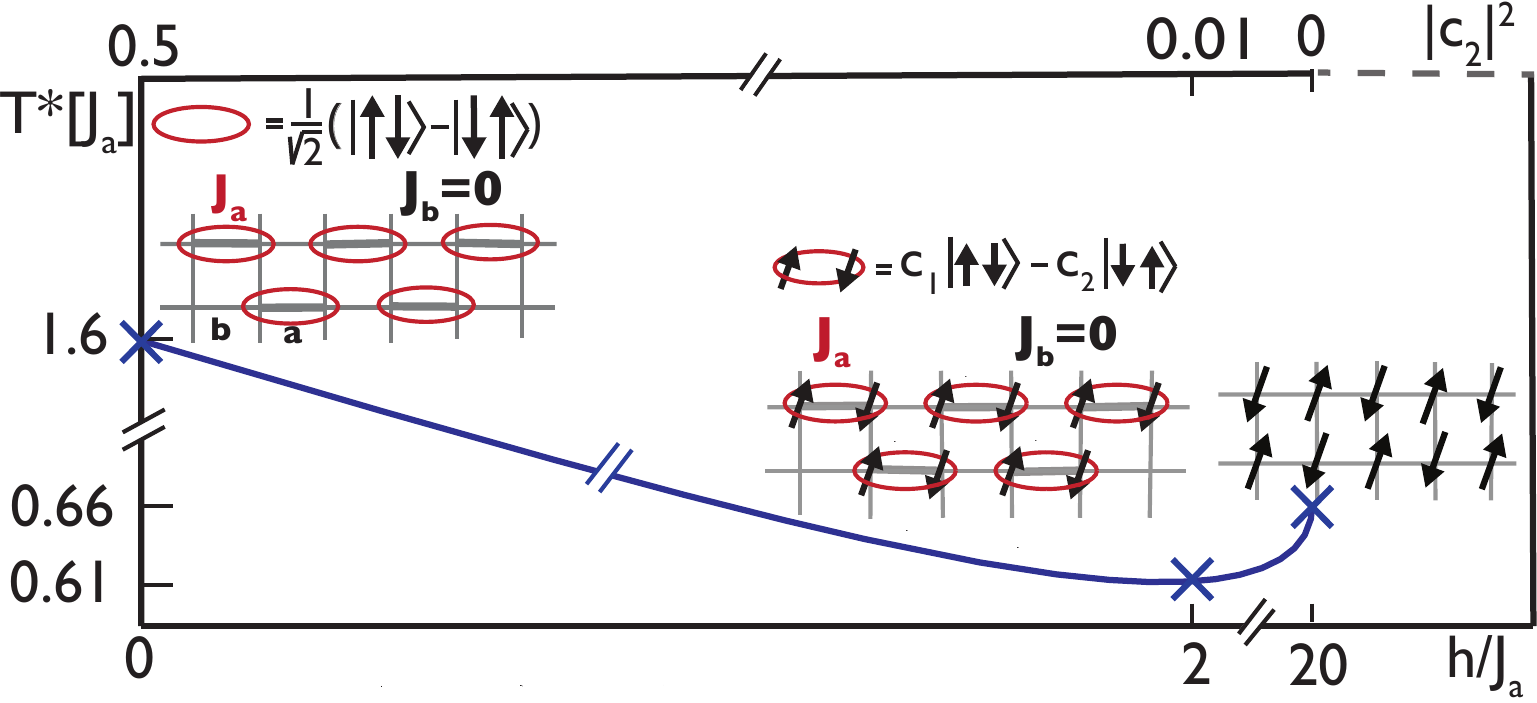} \label{figure1} \caption{\textit{(color online)} The effective temperature $T^{*}$ after the sudden quench as a function
of the initial state's coefficient $c_{2}$ or the magnetic field
$h$. Arrows represent spins, each red ellipse denotes a singlet state
\mbox{$(|\uparrow\downarrow\rangle-|\downarrow\uparrow\rangle)/\sqrt{2}$}
and ellipses surrounding arrows denote optimally polarized singlet
states \mbox{$0.99|\uparrow\downarrow\rangle-0.12|\downarrow\uparrow\rangle$}.
$J_{a}$ (bold) and $J_{b}$ (thin lines) are the intra- and inter-dimer
interactions.}
\end{figure}

How does $T^{*}$ depend on the size of the initial
state and on the speed of the quench?

To answer the first question, we consider a double well as the simplest
non-trivial subsystem, see Fig.~1. An array of double wells can be
created in a dimerized lattice \cite{Phillips2007,Bloch2008,Esslinger2013},
starting from a band insulating state and separating the sites employing
e.g. a frequency doubled optical lattice \cite{Bloch2008}. In our
limit of strong onsite interaction $U$, the atoms are treated as
Heisenberg spins with very weak inter-dimer interactions $J_{b}$
and strong intra-dimer coupling $J_{a}$. Naively, one might think
that the ideal choice for the initial state of the subsystem would
be a singlet state $|\psi\rangle=c_{1}|\uparrow\downarrow\rangle+c_{2}|\downarrow\uparrow\rangle$
with $|c_{1,2}|^{2}=1/2$, since this is the groundstate. However,
this state has zero staggered magnetization and hence only a small
overlap to the $T=0$ antiferromagnetic state. At the other extreme,
a classical N\'{e}el state with $|c_{2}|=0$ already contains a preferred
direction of broken symmetry and staggered magnetism. The drawback
here is the perfect ordering, which is not present in larger systems
even at $T=0$ due to quantum fluctuations. The optimal state is a
partially polarized state, with $|c_{2}/c_{1}|\approx8$, featuring
$T^{*}\approx0.61\, J_{a}$.

In the two-dimensional systems at finite temperature considered here,
only short range order with exponentially decaying correlations at
finite temperature exists. The correlation length $\xi(T)=A\exp{(2\pi\rho_{s}/k_{B}T)}$  
diverges exponentially for $T\rightarrow0$. Here $\rho_{S}=0.199$
is the renormalized spin-stiffness constant at $T=0$ and $A\approx0.276$
\cite{Ding1990,Manousakis1991,Chakravarty1989}. 
 However, the regime for which the correlation length $\xi(T)$ is larger than the lattice spacing $d$, is referred to as the antiferromagnetic (AF) regime. The crossover temperature
$T_{c}\approx0.97\, J$ is defined as $\xi(T_{c})=d$.
 As we show below, the optimal initial state of a product state of subsystems of size $L$ resembles the finite-temperature state in the following way: The optimal state of the subsystems 
  also has to have a nonzero magnetization, in analogy to the short-range order of the second-order phase transition,  and the optimal temperature $T^{*}$ approximately fulfills $\xi(T^{*}) \approx L$.
   The initial state therefore consists of ``patches'' of short-range order of length $L$, in analogy to the short-range ordered AF state with a correlation length $\xi(T)$. 
 This implies that even when increasingly larger
magnetically ordered subsystems with corresponding correlation lengths 
$\xi^{*}$ are prepared, the effective temperature $T^{*}$ obtained
after the quench only decreases logarithmically with $\xi^{*}$, as
we  show numerically. We note that  the correlation length $\xi(T)$ is experimentally visible in the broadening of the structure factor peaks that are obtained via Bragg scattering, in addition to technical broadening. 

Our scheme has two major advantages over related schemes: Firstly,
no significant redistribution of atoms is required \cite{Bernier2009,Ho2009}.
Such a redistribution is expected to cause severe problems due to
the extremely slow mass flow in the Mott insulating regime \cite{Chin2010}.
Secondly, the high entropy shell surrounding the central region of
interest does not necessarily have to be removed: In this case, the
timescale of the quench and the subsequent thermalization must be
shorter than the time for the entropy to leak back into the central
region. This condition will limit the correlation length, but at the
same time avoid the need for a removal step \cite{Bernier2009}, which
is technically very challenging and has not yet been demonstrated
experimentally. All ingredients necessary for the implementation of
this proposal have already been experimentally demonstrated \cite{Phillips2007,Bloch2008,Esslinger2013}. Merging plaquettes to create a state of fermionic pairs has been discussed in Ref. \cite{Rey2009}.  
Our scheme is closely related to the work by Lubasch and coworkers
\cite{Cirac2011}, where only double well subsystems initialized in
singlet states are considered. 

Throughout the paper, we consider fermions in the strongly repulsive
coupling regime $U\gg t$, where $t$ is the tunneling matrix element.
The system is hence well described by the two-dimensional Heisenberg
spin Hamiltonian 
\begin{equation}
H=J_{a}\sum_{{\langle ij\rangle}_{a}}\textbf{S}_{i}\textbf{S}_{j}+J_{b}\sum_{{\langle ij\rangle}_{b}}\textbf{S}_{i}\textbf{S}_{j}-\sum_{i}h_{i}S_{i}^{z},\label{HH}
\end{equation}
where $\langle ij\rangle_{a/b}$ means summation over nearest neighbours
along bonds $a$ and $b$, respectively, and $\textbf{S}_{i}$ is
the spin-$1/2$ operator. We set $\hbar\equiv1$ and $k_{B}\equiv1$.
The exchange couplings $J_{a/b}=4t_{a/b}^{2}/U>0$ quantify interactions
$J_{a}$ between spins in a subsystem and interactions $J_{b}$ between
subsystems. $t_{a/b}$ are the tunneling energies within/between subsystems. $h_{i}=(-1)^{i}h$ denotes the staggered magnetic field
pointing along the $z$ axis. We focus on staggered magnetic fields
with a spatially constant amplitude $h$ but also touch briefly on the
case with  a spatially inhomogeneous amplitude in section IV.A. Changing
the ratio $g\equiv J_{b}/J_{a}$,  the system undergoes a second order
phase transition. For columnar and for staggered dimerization it occurs
at $g_{c}=0.38$ \cite{Sachdev2000} and $g_{c}=0.396$ \cite{Jiang2013},
respectively. For $g>g_{c}$ the groundstate has N\'{e}el order, and for $T<T_{c}$ the system is in the so-called renormalized classical
regime.   For $g<g_{c}$ and for $T<\Delta$  
 the system in the quantum disordered region with a paramagnetic ground
state \cite{Chakravarty1989,Sachdev2000}, where $\Delta$ is the energy gap of the state.

We perform calculations for different initial subsystems such as double
wells, $2\times2$ plaquettes, $Z$-shaped and larger $n\times m$
subsystems, see Fig. 2b. The subsystems are always prepared in the
groundstate of (\ref{HH}) with $J_{b}=0$. To estimate the effective temperature, we assume that only low-energy modes are excited during the slow quench. Using linear spin-wave theory \cite{AA}
we analytically (numerically) estimate the temperature $T^{*}$ for
small (large) subsystems. We study how the effective temperature $T^{*}$
depends on the quench time $\tau$ and present analytical results
for slow changes.

\begin{figure}
\hfill{}\subfigure{\includegraphics[width=5.5cm]{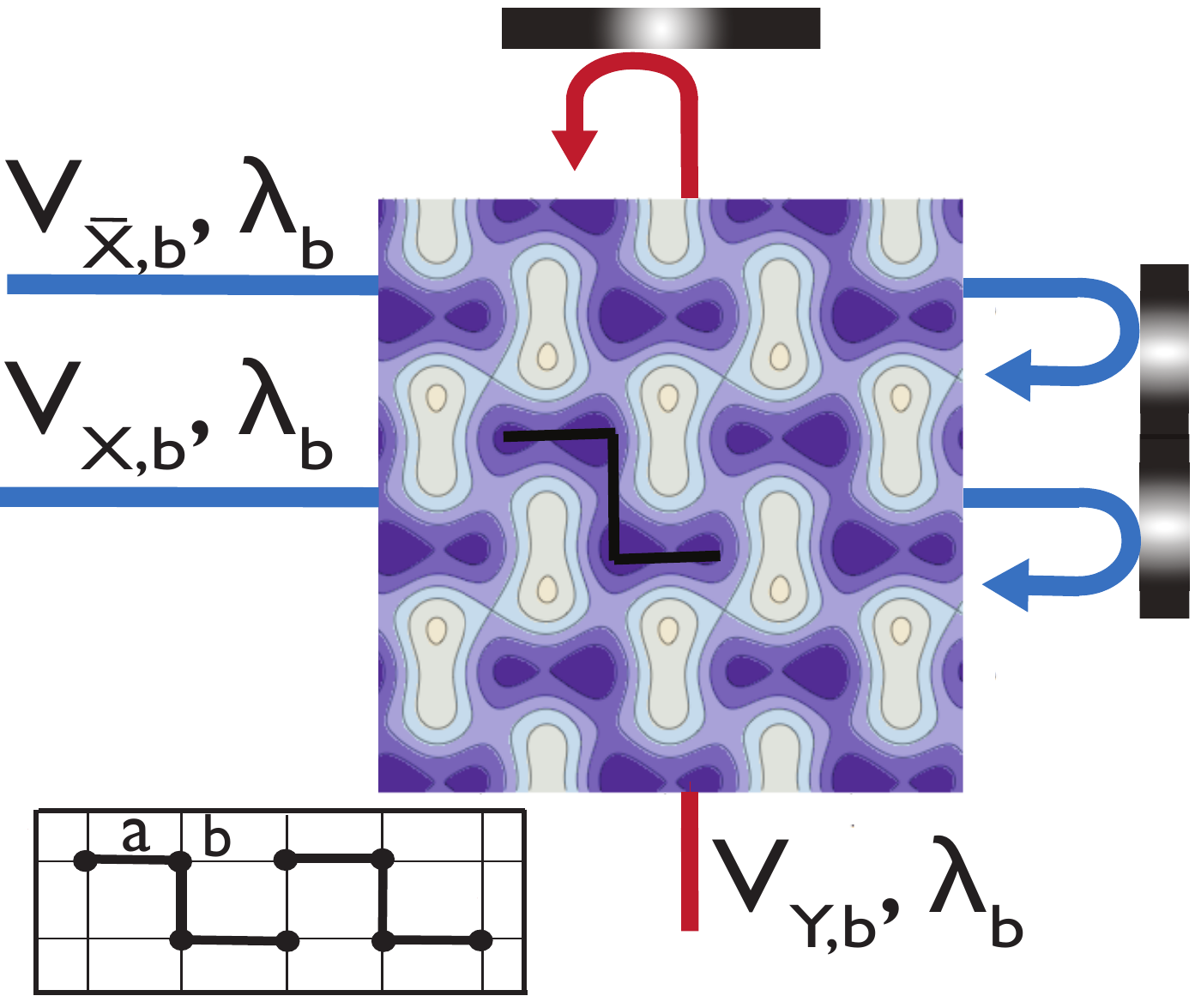}} \hfill{}\subfigure{\includegraphics[width=2.5cm]{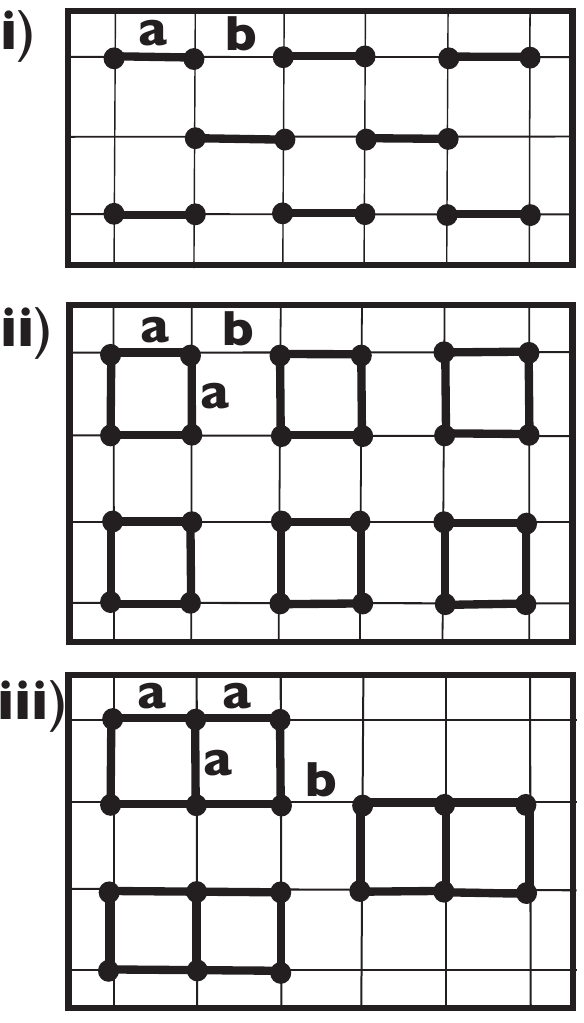}}
\hfill{}\caption{\textit{(color online)} (a) Contour plot of the Z-shape optical lattice. Beams $\bar{X}$,
$X$ propagate along the same axis and beam $Y$ along an axis perpendicular
to that axis. Beam $Y$ interferes with beam $X$. Lighter regions
correspond to higher potential energy and darker to lower energy.
We choose $V_{\bar{X},b}=V_{Y,b}=1,V_{X,b}=0.2$, $V_{X,r}=V_{Y,r}=0.3$
and $\phi_{1}=\pi/2,\,\,\phi_{2}=\pi/4.$ (b) Schematic pictures of
dimers (i), $2\times2$ plaquettes (ii) and $3\times2$ plaquettes
(iii).}
\end{figure}

\begin{center}
\textbf{II. LATTICE GEOMETRY} 
\par\end{center}
In our studies we consider several optical lattices, simple square
and dimerized lattices as well as plaquette- and $Z$-shaped lattices.
Dimerized and plaquette-shaped lattices have already been realised
experimentally using super-lattices \cite{Bloch2008,Bloch2012} or
a mixture of interfering and non-interfering lattices \cite{Phillips2007,Esslinger2013}.

Here we describe a technique to create an array of four-site states
in $Z$-shape configuration, see Fig. 2a. This is the most complicated
geometry which we still deem realizable with current experimental
techniques. Due to the $Z$-shape, a staggered magnetic field can be
created employing only a magnetic field gradient pointing along the
diagonal axis. The larger $n\times m$ plaquettes depicted in Fig.
2b-iii serve as theoretical constructs to discuss scaling issues.

For each lattice, the fundamental building block is a repulsive cubic
optical lattice $V_{\bar{X},Y}(x,y)$ generated by two blue-detuned
retro-reflected beams $\bar{X}$ and $Y$, as depicted in Fig. 2a.
The two beams have a wavelength of e.g. $\lambda_{b}=532$ nm and
a slight frequency offset to avoid mutual interference. In order to
make effective bonds along the $x$-direction, an additional retro-reflected
beam $X$ along the $x$-direction is used, which interferes with
beam $Y$. The resulting potential is $V_{X,Y}(x,y)$. The bonds along
the $y$-direction are created by an attractive cubic optical lattice
$V_{\lambda_{r}}(x,y)$ formed by two red-detuned retro-reflected
beams at a wavelength $\lambda_{r}=2\lambda_{b}$ (beams not shown
in Fig. 2a). The resulting trapping potential is presented in the
Fig. 2a and given by the equation 
\begin{eqnarray}
V(x,y) & = & V_{\bar{X},Y}(x,y)+V_{X,Y}(x,y)+V_{\lambda_{r}}(x,y)\\
 & = & +V_{\bar{X},b}\cdot\cos^{2}(k_{b}x)+V_{Y,b}\cdot\cos^{2}(k_{b}y)\nonumber \\
 &  & +V_{X,b}\cdot\cos^{2}(k_{b}x+\phi_{1})\nonumber \\
 &  & +2\sqrt{V_{X,b}V_{Y,b}}\cdot\cos(k_{b}x+\phi_{1})\cos(k_{b}y)\nonumber \\
 &  & -V_{X,r}\cdot\cos^{2}(k_{r}x)-V_{Y,r}\cdot\cos^{2}(k_{r}y+\phi_{2}),\nonumber 
\end{eqnarray}
where $V_{\bar{X},b},V_{X,b},V_{Y,b}$ and $V_{X,r},V_{Y,r}$ are
the respective lattice depths. They are given in units of the recoil
energy $E_{b}=h^{2}/2m\lambda_{b}^{2}$ for $V_{\bar{X},b},V_{X,b},V_{Y,b}$
and $E_{r}=h^{2}/2m\lambda_{b}^{2}$ for $V_{X,r},V_{Y,r}$. $m$
denotes the mass of a single atom.\\

\noindent \begin{center}
\textbf{III. SPIN WAVE THEORY} 
\par\end{center}
To describe the final state after the quench and estimate $T^{*}$,
we use linear spin-wave theory following Holstein and Primakoff (HP) \cite{Holstein1940}, in the notation of Ref. \cite{AA}. 
 We choose the classical N\'{e}el state along the $z$ and $-z$ direction and define the rotated spin
  $\tilde{\bf S}_{i} \equiv (S_{i}^{x}, -S_{i}^{y}, -S_{i}^{z})$, for sites $i$ on one sublattice. 
 Then the Hamiltonian reads
\begin{equation}
H=-|J|\sum_{\langle ij\rangle}S_{i}^{z}\tilde{S}_{j}^{z}+\frac{|J|}{2}\sum_{\langle ij\rangle}(S_{i}^{+}\tilde{S}_{j}^{+}+S_{j}^{-}\tilde{S}_{i}^{-}).\label{ham1}
\end{equation}
 where we note that after the quench the spin couplings 
are equal, i.e., $J_{a}=J_{b}\equiv J$.
 We now  approximate the spins by bosonic operators 
\begin{align}
 & S_{i}^{+}=\sqrt{2S - n_{i}}a_{i},\, S_{i}^{-}=a_{i}^{\dagger} \sqrt{2S- n_{i}},\, 
  S_{i}^{z}=S-n_{i},\label{HP}
\end{align}
 and similarly for $\tilde{\bf S}_{i}$. 
 The operators $a_{i}$, $a_{i}^{\dagger}$
satisfy commutation relations $\left[a_{i},a_{j}^{\dagger}\right]=\delta_{ij}$, and 
 $n_{i} \equiv a_{i}^{\dagger}a_{i}$ is the number of bosons on site $i$. It is limited by $2S$, with $S=1/2$. We expand the square root in
 Eqs. (\ref{HP}) as $\sqrt{1 - n_{i}/2S} \approx1-\frac{n_{i}}{4S}-\cdots$
and carry out the calculations to the lowest order, i.e., perform the
linear spin wave approximation. This leads to an ensemble of non-interacting
bosonic modes.

\begin{figure*}
\includegraphics[scale=0.74]{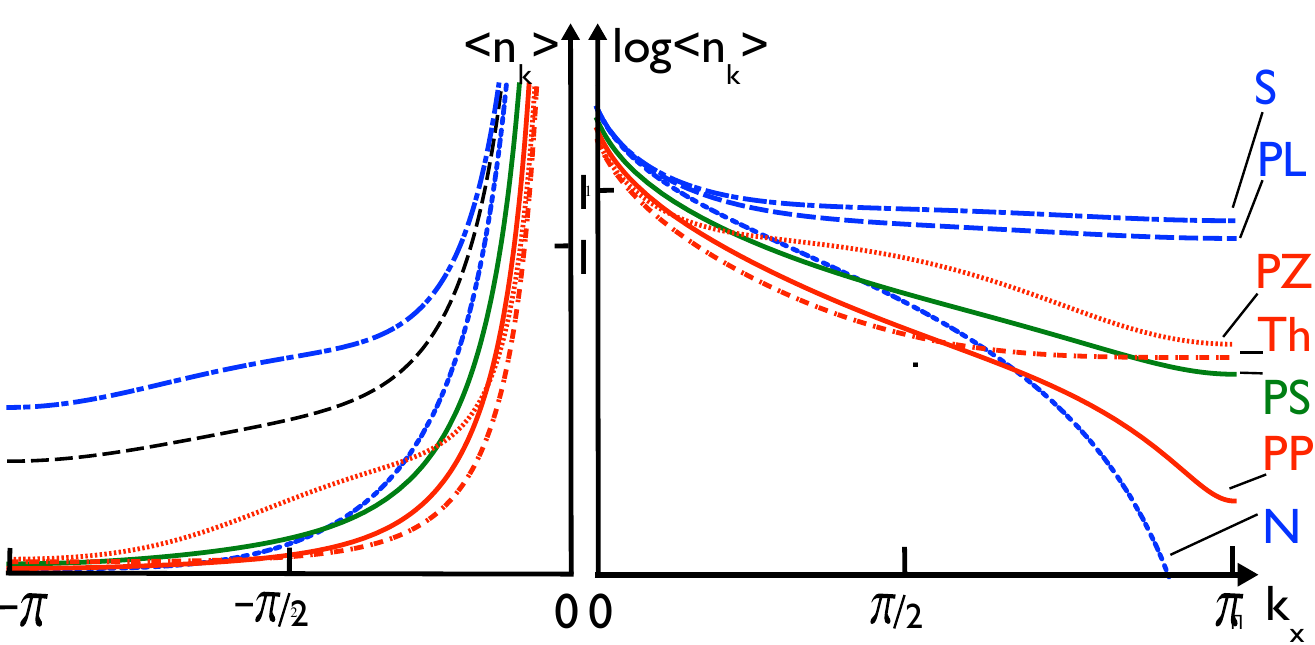} \caption{\textit{(color online)} The average number of excitations $\langle n_{\textbf{k}}\rangle$
 on a linear and logarithmic scale as a function of $k_{x}$, for $k_{y}=0$.
The initial states are: N\'{e}el (N), singlet (S), polarized singlet (PS),
plaquette (PL), polarized plaquette (PP), polarized $Z$-state (PZ)
and a thermal distribution (Th) with $T=0.554\, J$. For the polarized
cases the optimal magnetic field yielding the lowest effective temperature
was chosen.}
\label{plotnk} 
\end{figure*}

We first insert the linearized bosonic operators of Eq. (\ref{HP})
into the Hamiltonian (\ref{ham1})
 and  use the Fourier transform 
 $a_{\textbf{k}}=\frac{1}{\sqrt{N}}\sum_{i}e^{-i\textbf{k}\textbf{r}_{i}}a_{i}$
where $\textbf{k}$ runs over the first Brillouin zone, $\textbf{k}\in[-\pi/a,\pi/a)^{2}$, where 
$a$ is the lattice constant, set to $a\equiv1$. $N$ is the number of lattice sites.
 The Hamiltonian becomes
\[
H=-\frac{NzJS^{2}}{2}+zJS\sum_{\textbf{k}}(a_{\textbf{k}}^{\dagger}a_{\textbf{k}}+\frac{\gamma_{\textbf{k}}}{2}(a_{\textbf{k}}a_{-\textbf{k}}+a_{\textbf{k}}^{\dagger}a_{-\textbf{k}}^{\dagger})),
\]
where $\gamma_{\textbf{k}}\equiv \frac{1}{2}(\cos k_{x}+\cos k_{y})$, 
 and $\textbf{k}=(k_{x},k_{y})$.
$z$ is the number of the
nearest neighbours, $z=2d$, and $d=2$ is the lattice dimensionality.
 Then we perform a Bogoliubov
transformation $a_{\textbf{k}}=u_{\textbf{k}}\alpha_{\textbf{k}}-v_{\textbf{k}}\alpha_{-\textbf{k}}^{\dagger}$,
with $u_{k}=\cosh\theta_{k}$ and $v_{k}=\sinh\theta_{k}$, where
the $\alpha_{\textbf{k}}$ are bosonic operators. To cancel anomalous
 terms in the Hamiltonian we choose $\tanh2\theta_{\textbf{k}}=-\gamma_{\textbf{k}}$.
The Hamiltonian now has the desired diagonal form 
\begin{equation}
H_{LSW}=-\frac{NzJS^{2}}{2}+\sum_{\textbf{k}}\omega_{\textbf{k}}(\alpha_{\textbf{k}}^{\dagger}\alpha_{\textbf{k}}+\frac{1}{2}),\label{lsw}
\end{equation}
where $\omega_{\textbf{k}}=JSz\sqrt{1-\gamma_{\textbf{k}}^{2}}$ is
the spin-wave dispersion relation. Near $\textbf{k}=\lbrace0,0\rbrace$
and $\textbf{k}=\lbrace\pi,\pi\rbrace$ the dispersion is linear with
$\omega_{\textbf{k}}\approx JSz\mid\textbf{k}\mid$ and $\omega_{\textbf{k}}\approx JSz\mid\textbf{k}-(\pi,\pi)\mid$,
respectively. These two Goldstone modes reflect the broken symmetry
of the antiferromagnetic state. Due to the absence of a gap the creation
of excitations cannot be avoided even for a slow quench. The staggered
magnetization $M=\frac{1}{N}\sum_{i}(-1)^{i}\langle S_{i}^{z}\rangle$
for spin-$1/2$ in the linear spin-wave approximation is $M\approx0.303$
\cite{Manousakis1991, AA}, so quantum fluctuations lead to a reduction
 to about 60$\%$ of the classical value. Numerical and
theoretical studies of the staggered magnetization in the infinite-lattice
extrapolation for spin-$1/2$ Heisenberg models have found that $M\sim0.25-0.4$
\cite{Manousakis1991}. \\

\begin{center}
\textbf{IV. SUDDEN QUENCH} 
\par\end{center}
We now describe the method to determine the effective temperature
$T^{*}$ after the quench. In short, we calculate the momentum distribution
of spin waves created in the quench using linear spin-wave theory,
$\langle n_{\textbf{k}}\rangle=\sum_{i,j}c_{i}^{*}c_{j}\langle i|\alpha_{\textbf{k}}^{\dagger}\alpha_{\textbf{k}}|j\rangle$,
where the summation is over the basis states $| i \rangle$ of the subsystem, and $|\Psi\rangle \equiv \sum_{i} c_{i} |i\rangle$ is the groundstate. The
average number of excitations $\langle n_{\textbf{k}}\rangle$ as
a function of $k_{x}$ (and $k_{y}=0$) for various geometries is
plotted in Fig. \ref{plotnk}. The energy of the system, compared to the groundstate, after the
quench is given by $\langle E\text{\ensuremath{\rangle}=}\sum_{\mathbf{k}}\langle n_{\textbf{k}}\rangle\omega_{k}$.
Since the system is isolated, the energy after the quench is conserved.
The effective temperature $T^{*}$ is then defined to be the temperature
of a thermal distribution having the same energy $\langle E^{th}\text{\ensuremath{\left(T^{*}\right)\rangle}=}\sum_{\mathbf{k}}\langle n_{\textbf{k}}^{th}(T^{*})\rangle\omega_{k}$.
An alternative measure for the temperature scale $T_{k}^{*}$ can
be defined by finding the thermal distribution $\langle n_{\textbf{k}}^{th}(T_{k}^{*})\rangle$
which matches the post-quench distribution $\langle n_{\textbf{k}}\rangle$
at small momenta, yielding $T_{k}^{*}=\lim_{k\rightarrow0}\langle n_{\textbf{k}}\rangle\omega_{k}$.
  We note that this limit is well-defined, because for the cases we discuss here, the limit is independent of how ${\bf k }$ approaches zero.
 We use the former definition $T^{*}$, since it takes into account
not only low energy but rather all excitations created. We note that
$T_{k}^{*}$ appears when we consider slow quenches in section VI.

\begin{table}
\includegraphics[scale=0.9]{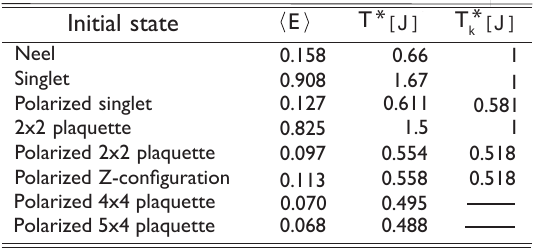}\caption{TAB. I. The minimal energy $\langle E\rangle$ and the minimal tem\-peratures
$T^{*}$ and $T_{k}^{*}$ at optimal polarization. For effective temperatures $T^{*}<T_{c}=0.97\, J$ \cite{Chakravarty1989,Manousakis1991,Ding1990}
antiferromagnetic ordering is expected.}
\label{TabI} 
\end{table}

\begin{center}
\textbf{A. Homogeneous magnetic field} 
\par\end{center}

The precise protocol of the quench is as follows: Initially, we assume
that the coupling between subsystems is $J_{b}=0$, and the amplitude
of the staggered magnetic field $h$ can chosen freely. The quench
is initiated by suddenly  increasing the coupling between subsystems,
i.e., setting $J_{b}=J_{a}$. At the same time the magnetic field is
turned off. The groundstate before the quench $|\Psi\rangle$ is
found by solving the Heisenberg Hamiltonian in eq. (\ref{HH}) in
linear spin wave approximation, see eq. (\ref{lsw}). The momentum
distribution is determined by evaluating $\langle\Psi|\alpha_{\textbf{k}}^{\dagger}\alpha_{\textbf{k}}|\Psi\rangle$.
We perform analytical calculations for the N\'{e}el state, singlet state,
$2\times2$ plaquette and $Z$-shape state. Details of the calculations
are presented in the Appendix. For the larger $n\text{\ensuremath{\times}}m$
plaquettes we carry out numerical calculations using exact diagonalization.
Table I summarizes the effective temperatures found for the different
configurations. We observe that with the proposed scheme the regime
of antiferromagnetic correlations can be easily reached, with e.g.
$T^{*}\approx0.63\, T_{c}$ for optimally polarized singlets. There
is good agreement between the two effective temperatures for optimally
polarized states, which minimize high-energy excitations. The much
larger discrepancy for unpolarized states stems from the fact that
only the low energy part of the distribution is used to determine
$T_{k}^{*}$.

Fig. 4 depicts numerical results for systems up to 20 spins in a $5\times4$
configuration. As the magnetic field increases, the effective temperature
of the system decreases up to the minimum value $T^{*}$. After exceeding
the optimal value of the magnetic field all curves tend towards the
temperature of the classical N\'{e}el state $T^{*}\approx0.66\, J$. The
plot shows that the magnetization and the effective temperature go
to zero with increasing size of the subsystem. We find that $T^{*}$
depends weakly on the subsystem size, possibly indicating logarithmic
scaling.

\begin{figure*}
\includegraphics[scale=2.0]{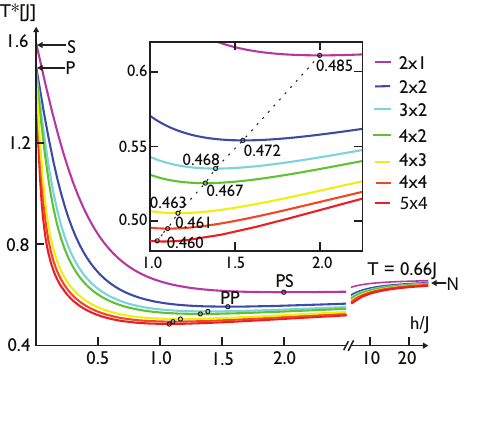} \caption{\textit{(color online)} The effective temperature $T^{*}$ as a function of the magnetic field
for larger plaquette subsystems consisting of $m\times n$ sites.
The inset shows the regime of $h/J_{a}\in[1.0-2.5]$. The circles
mark the minimal effective temperature for each lattice geometry and
the labels indicate the corresponding staggered magnetization $M=\frac{1}{N}\sum_{i}\langle S_{i}^{z}\rangle$.
It should approach the groundstate value $M\approx0.303$ of the
linear spin-wave approximation as the system size goes to infinity.
The dotted line is a guide to the eye. The configuration labels conform
to the nomenclature in Fig. \ref{plotnk}. }
\end{figure*}

The explanation of this behaviour lies in the dependence of the correlation
length on the temperature $\xi(T^{*})$. Inverting this formula
and assuming that the correlation length is comparable to the linear
size of a subsystem yields $T^{*}=\frac{2\pi\rho_{S}}{\log{(L/A)}}$, where 
 we choose $L = \sqrt{m\cdot n}$ as the length scale of the subsystem. 
 The curve is plotted in Fig. 5a and compared to our numerical
results (dots). We note that a $2\times2$ plaquette and the $Z$-system
give almost the same energy $T^{*}$. This is consistent with the
hypothesis that $T^{*}$ scales only with the linear system size $L$.

\begin{figure}
\hfill{}\subfigure{\includegraphics[width=6.0cm]{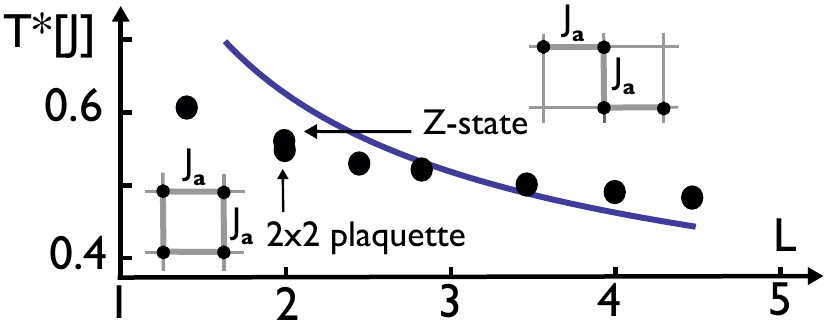}} \hfill{}\subfigure{\includegraphics[width=2.5cm]{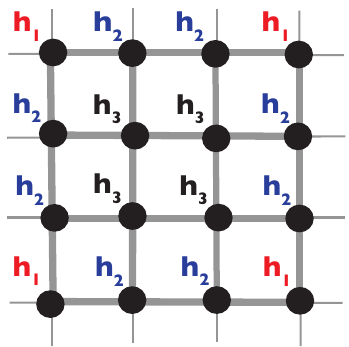}}
\hfill{}\caption{\textit{(color online)} (a) The minimal temperature $T^{*}$ as a function of the linear size
of subsystems $L$. There is reasonable agreement between the numerical
results (dots) with the expected logarithmic scaling (continuous line)
expected from theory with $\rho_{S}=0.199$ and $A=0.276$ \cite{Ding1990}.
(b) Distribution of inhomogenous staggered magnetic fields over a
$4\times4$ plaquette, the field values $h_{1},h_{2},h_{3}$ characterize
the distribution.}
\end{figure}

\begin{center}
\textbf{B. Inhomogeneous magnetic field} 
\par\end{center}
So far we only treated the polarization of the initial state by a
staggered magnetic field of constant amplitude $h$. Now we take into
account a site-dependent staggered magnetic field. We consider only
the case of a $4\times4$ plaquette. The magnetic field on each plaquette
is distributed in the way shown in Fig. 5b, with alternating polarity
on adjacent sites. The total energy $\langle E\rangle$ is minimized
by varying the three parameters $h_{1},h_{2},h_{3}$. We find that
the lowest energy corresponds to the case where the outer sites of
the plaquette are subjected to a higher magnetic field than those
in the center. The energy reaches its minimum value $T_{inh}^{*}$
for $h_{1}/J=1,h_{2}/J=0.7,h_{3}/J=0.45$. However, the reduction
in effective temperature with respect to the $4\times4$ plaquette
in a homogeneous staggered field by only $0.003\, J$ to $T_{inh}^{*}=0.492\, J$
is rather small.

\begin{center}
\textbf{VI. SLOW QUENCH} 
\par\end{center}
In this section we discuss the effect of a finite quench time $\tau$
on the effective temperature. We consider N\'{e}el, singlet, optimally
polarized singlet states and optimally polarized $2\times2$ plaquettes
in a homogeneous staggered magnetic field. The system is prepared
in the same way as in the sudden quench, but the interaction $J_{b}$
between subsystems is turned on within a time interval $\tau$, and the magnetic field $h$ is turned off smoothly on the same time scale. To
estimate the effective temperature we make the simplifying assumption
that modes with frequency $\omega_{\textbf{k}}>1/\tau$ adapt adiabatically
to the parameter changes whereas modes with $\omega_{\textbf{k}}<1/\tau$
are populated as in a sudden quench. The total energy after the quench
can then be calculated as $\langle E(\tau)\rangle=\sum_{\textbf{k}}\omega_{k}\langle n_{k}\rangle\Theta(\omega_{\textbf{k}}-1/\tau)$.
As before, we find $T^{*}$ by finding the thermal distribution having
an energy $\langle E(\tau)\rangle$. As shown in Fig. 6, the effective
temperature drops rapidly with increasing quench time. Moreover, the
effective temperatures for N\'{e}el and  singlet case 
approach each other since their excitation densities at small momenta approach the same value.

\begin{figure}
\includegraphics[width=80mm]{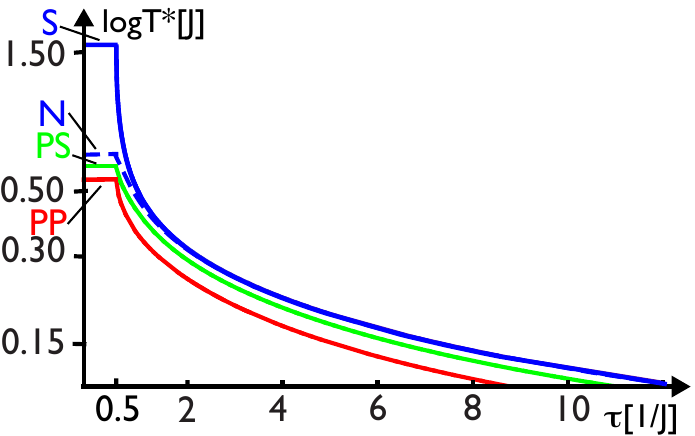} \caption{\textit{(color online)} The effective temperature $T^{*}$ as a function of the quench time
$\tau$ on a log-lin scale. The different configurations are labeled
as in previous figures. As $\tau$ increases, the effective temperature $T^{*}$ quickly approaches power-law scaling,  $T^{*}\sim(T_{k}^{*}/\tau^{2})^{1/3}$.}
\end{figure}

It is instructive to find the scaling of the effective temperature
with quench time in the limit of  slow quenches. As it turns out,
the effective temperature $T^{*}$ is related to the temperature scale
$T_{k}^{*}$ associated with the low-frequency excitations and the
sweep time according to $T^{*}\propto\left(\frac{T_{k}^{*}}{\tau^{2}}\right)^{1/3}$.
To show this, we consider the limit in which only modes with $\textbf{k}\approx0$
are occupied, which have a linear dispersion $\omega_{\textbf{k}}\sim\vert\textbf{k}\vert$
and consequently an excitation probability $\langle n_{\textbf{k}}\rangle\approx T_{k}^{*}/\omega_{\textbf{k}}$.
This gives the following dependence of the total energy on $\tau$:
\begin{equation}
\langle E^{ad}(\tau)\rangle=\frac{1}{(2\pi)^{2}}\int\limits _{\omega_{\textbf{k}<1/\tau}}d^{2}\textbf{k}\,\omega_{k}\langle n_{\textbf{k}}\rangle=\frac{1}{4\pi}\frac{T_{k}^{*}}{\tau^{2}}.
\end{equation}
We can compare this energy to the thermal energy for small temperatures:
\begin{equation}
\langle E_{T\rightarrow0}^{th}\rangle=\frac{1}{(2\pi)^{2}}\int\limits _{0}^{\textbf{k}_{cut}}d^{2}\textbf{k}\,\omega_{k}\langle n_{\textbf{k}}^{th}\rangle=\zeta(3)\frac{T^{*^{3}}}{\pi},
\end{equation}
where we again use that for $k\rightarrow0$ the spin wave spectrum
is linear and $\zeta(3)\approx1.2$ is the Riemann zeta function.
Equating these energies we find that for  slow quenches the effective
temperature changes according to $T^{*}=\left(\frac{1}{4\zeta(3)}\frac{T_{k}^{*}}{\tau^{2}}\right)^{1/3}$.
 In this limit, the figure of merit to be optimized is thus $T^{*}_{k}$. 
  We note that for even slower quenches, this power law behavior will eventually be controlled by the critical point at $g_{c}$. In this limit,  the power-law exponent would be replaced by  critical exponents. 

\begin{center}
\textbf{VII. SUMMARY} 
\par\end{center}
In conclusion, we have proposed a method and a specific experimental realization to reach the antiferromagnetically
ordered state by preparing very low entropy subsystems, which do not
interact initially. The subsystems are merged by either sudden or
slow quenches of the interaction, where slower quenches result in significantly 
lower effective temperatures. The effective temperature is calculated
within linear spin-wave theory and we observe that it reduces with increasing linear subsystems size. 
Assuming that the subsystem size determines the correlation length of the resulting
antiferromagnetic state, we expect logarithmic scaling. 
 Additionally,  we find that the effective temperature can be 
reduced if the subsystems are prepared with optimal polarization,
closely resembling the target state. We expect that these insights will
be very useful to attain antiferromagnetic ordering in experiments
employing double well or plaquettes type geometries. 
 As shown in Sect. II, the method can be implemented with current technology. 
 Furthermore, the principle that is outlined here, i.e., to assemble subsystems
    with precursors of the desired short-range order, can also be applied to the construction
     of other many-body states in ultracold atom systems.

We acknowledge support from the Deutsche Forschungsgemeinschaft through
the SFB 925 and the Landesexzellenzinitiative Hamburg, which is supported
by the Joachim Herz Stiftung.\newpage{}

\begin{center}
\textbf{APPENDIX} 
\par\end{center}

\begin{appendices} In our analytical computations we consider the
following states:

\begin{enumerate}
\item Classical N\'{e}el state. This state is a product state of spins up $\vert\uparrow\rangle$
and spins down $\vert\downarrow\rangle$ ordered in a checkerboard
$\vert0\rangle\equiv|\uparrow\downarrow\uparrow\downarrow...\rangle$. 
\item Polarized singlet state. This is a product state of singlets which
are arranged as shown in picture Fig.2b i. We represent it by the
formula: \\
 $\Big(\prod_{\text{i}}(c_{1}-c_{2}\alpha_{i,1}^{\dagger}\alpha_{i,2}^{\dagger})\Big)|0\rangle$.\\
 In the presence of nonzero $h$, the coefficients $c_{1},c_{2}$ are
generally not equal to each other only for $h=0$, $c_{1}=c_{2}$, 
and the standard singlet state is recovered by putting $c_{1}=1/\sqrt{2}$.\\

\item Polarized 2$\times$2 plaquette state. This is a product state of
$2\times2$ plaquettes which are arranged as shown in picture Fig.2b
ii. We represent it by the formula:\\
 $\Big(\prod_{i}(c_{1}+c_{2}\prod_{j=1}^{4}\alpha_{i,j}^{\dagger}+c_{3}\sum_{j=1}^{4}\alpha_{i,j}^{\dagger}\alpha_{i,j+1}^{\dagger})\Big)|0\rangle$.
For $h=0$ the coefficients are following $c_{1}=c_{2}=-\frac{1}{\sqrt{3}},c_{3}=\frac{1}{2\sqrt{3}}$.\\

\item Polarized $Z$-state. This is a product state of $Z$-states which
are arranged as shown in picture Fig.2a. We represent it by the formula:\\
 $\Big(\prod_{i}(c_{1}+c_{2}\prod_{j=1}^{4}\alpha_{i,j}^{\dagger}+c_{3}\alpha_{i,2}^{\dagger}\alpha_{i,3}^{\dagger}+c_{4}\alpha_{i,1}^{\dagger}\alpha_{i,4}^{\dagger}+c_{5}(\alpha_{i,1}^{\dagger}\alpha_{i,2}^{\dagger}+\alpha_{i,3}^{\dagger}\alpha_{i,4}^{\dagger}))\Big)\vert0\rangle$ 
\end{enumerate}
\noindent In all above states $|0\rangle$ is the classical N\'{e}el state
, $\alpha_{i}$, $\alpha_{i}^{\dagger}$ are the annihilation and
creation bosonic quasiparticle operators at the lattice site $i$.
They are Fourier transformed operators discussed before in the part
'Spin wave theory.' All states are normalised, $\sum_{i}|c_{i}|^{2}=1$.\\
 \indent The momentum distributions of Bogoliubov excitations created
in the sudden quench for the above states are following:

\begin{enumerate}
\item For N\'{e}el state, 
\begin{equation}
\langle n_{\textbf{k}}\rangle=\frac{1}{2\sqrt{1-\gamma_{\textbf{k}}^{2}}}-\frac{1}{2},
\end{equation}

\item For polarized singlet state 
\begin{align}
 & \left\langle n_{\textbf{k}}\right\rangle =\frac{c_{2}^{2}+\frac{1}{2}-\gamma_{\textbf{k}}c_{2}\sqrt{1-c_{2}^{2}}\cos{k_{x}}}{\sqrt{1-\gamma_{\textbf{k}}^{2}}}-\frac{1}{2}
\end{align}

\item For polarized 2$\times$2 plaquette state 
\begin{align}
 & \left\langle n_{\textbf{k}}\right\rangle =\frac{f_{1}(\textbf{c})}{2\sqrt{1-\gamma_{\textbf{k}}^{2}}}-\frac{1}{2}+2c_{3}^{2}\sin{k_{x}}\sin{k_{y}},
\end{align}
where $\textbf{c}=(c_{1},c_{2},c_{3})$ and 
\begin{eqnarray}
f_{1}(\textbf{c}) & = & 2c_{2}^{2}+4c_{3}^{2}(1+\cos k_{x}\cos k_{y})+\\
 &  & (\cos k_{x}+\cos k_{y})^{2}(c_{2}c_{3}+c_{1}c_{3})+1,\nonumber 
\end{eqnarray}

\item For polarized $Z$-state,\\

\begin{equation}
\left\langle n_{\textbf{k}}\right\rangle =\frac{f_{2}(\textbf{c})}{2\sqrt{1-\gamma_{\textbf{k}}^{2}}}-\frac{1}{2},
\end{equation}
where $\textbf{c}=(c_{1},...,c_{5})$ and $f_{2}(\textbf{c})=\sum_{i=1}^{5}C_{i}(\textbf{c})$\\

\begin{eqnarray}
C_{1}(\textbf{c}) & = & 1+2(c_{2}^{2}+c_{5}^{2})+c_{3}^{2}+c_{4}^{2}\nonumber \\
C_{2}(\textbf{c}) & = & 2(c_{3}+c_{4})B_{5}\cos{(k_{x}-k_{y})}\nonumber \\
C_{3}(\textbf{c}) & = & 2\gamma_{\textbf{k}}(c_{1}+c_{2})c_{5}\cos k_{x}\\
C_{4}(\textbf{c}) & = & \gamma_{\textbf{k}}(c_{2}c_{3}+c_{1}c_{5})\cos{(2k_{x}-k_{y})}\nonumber \\
C_{5}(\textbf{c}) & = & \gamma_{\textbf{k}}(c_{1}c_{3}+c_{2}c_{5})\cos k_{y}.\nonumber 
\end{eqnarray}

\end{enumerate}
In all above formulas $\gamma_{\textbf{k}}=\frac{1}{2}(\cos{k_{x}}+\cos{k_{y}})$.\\

\indent Now we calculate the total energy after the sudden quench,
$\langle E\rangle=\sum_{\textbf{k}}\omega_{\textbf{k}}\langle n_{\textbf{k}}\rangle$
in all above states, $\omega_{\textbf{k}}$ is the spin wave dispersion.
Energies in the thermodynamic limit are the following:
\begin{enumerate}
\item For N\'{e}el state, $\langle E\rangle\approx0.158$. 
\item For polarized singlet state, 
\begin{equation}
\langle E\rangle\approx2c_{2}^{2}+\frac{1}{2}c_{2}\sqrt{1-c_{2}^{2}}+0.16.
\end{equation}

\item For polarized 2$\times$2 plaquette state, 
\begin{equation}
\langle E\rangle\approx2(c_{2}^{2}+2c_{3}^{2})+(c_{2}c_{3}+c_{1}c_{3})+0.16.
\end{equation}

\item \noindent For polarized $Z$-state 
\begin{align}
 & \langle E\rangle\approx2(c_{2}^{2}+c_{5}^{2})+c_{3}^{2}+c_{4}^{2}+\\
 & \frac{1}{4}(c_{1}c_{3}+c_{2}c_{4}+2c_{1}+2c_{2})+0.16.\nonumber 
\end{align}

\end{enumerate}
\indent Above energies are convex functions of $c_{i}$'s so they
can be optimized over the external magnetic field $h$. The Lagrangian
multipliers are used to estimate their minima. Energies with optimal
magnetic field $h$ are compared to the energy of thermal distribution
with temperature $T^{*}$ in the thermodynamic limit, $\langle E^{th}\rangle=\frac{1}{4\pi^{2}}\int d^{2}\textbf{k}\,\omega_{\textbf{k}}\langle n_{\textbf{k}}^{th}(T^{*})\rangle$.
The resulting temperature is denoted as $T^{*}$.\\
 \indent Now we approximate the effective temperature $T^{*}$ for
an adiabatic quench. In this case we calculate the number of spin
waves in the long-wavelength limit $\langle n_{{k_{x}}}\rangle\equiv\langle n_{\textbf{k}}\rangle_{k_{x}\rightarrow0}$ (we set $k_{y}=0$):

\begin{enumerate}
\item For N\'{e}el state, $\langle n_{k_{x}}\rangle=\frac{1}{2\sqrt{1-\gamma_{\textbf{k}}^{2}}}$. 
\item For polarized singlet state, $\langle n_{k_{x}}\rangle=\frac{g_{1}(c_{2})}{\sqrt{1-\gamma_{\textbf{k}}^{2}}},$
where $g_{1}(c_{2})=c_{2}^{2}+\frac{1}{2}-c_{2}\sqrt{1-c_{2}^{2}}$. 
\item \mbox{For polarized 2$\times$2 plaquette state, $\left\langle n_{k_{x}}\right\rangle =\frac{g_{2}(\textbf{c})}{\sqrt{1-\gamma_{k}^{2}}},$}
where $\textbf{c}=(c_{1},c_{2},c_{3})$ and 
\begin{equation}
g_{2}(\textbf{c})=(c_{2}^{2}+2c_{3}^{2})+2c_{3}^{2}+2(c_{2}c_{3}+c_{1}c_{3})+\frac{1}{2}.
\end{equation}

\item For polarized $Z$-state, $\left\langle n_{k_{x}}\right\rangle =\frac{g_{3}(\textbf{c})}{\sqrt{1-\gamma_{k}^{2}}}$,\\
 where $\textbf{c}=(c_{1},...,c_{5})$ and 
\begin{equation}
g_{3}(\textbf{c})=\frac{1}{2}(2c_{2}^{2}+\sum_{i=3}^{5}c_{i}^{2})+\frac{1}{2}((c_{1}+c_{2})(c_{3}+c_{4}))+c_{5}\sum_{i=1}^{4}c_{i}+\frac{1}{2}.
\end{equation}

\end{enumerate}
Since $g_{i}(\textbf{c})$ are convex functions, there is a magnetic
field which optimizes $\langle n_{k_{x}}\rangle$. We use again the
Lagrangian multipliers to find minima of $\langle n_{k_{x}}\rangle$.
We notice that $\langle n_{k_{x}}\rangle$ diverges the same way as a 
thermal distribution for small energies $\langle n_{\textbf{k}}^{th}\rangle\approx\frac{T^{*}}{\omega_{k}}$.
Comparing them we obtain the minimal temperature $T_{k_{x}}^{*}$. \end{appendices}

\end{document}